\documentclass[a4paper]{article}

\usepackage{itgspeech2023}    %% Include ITGSpeech2023 style
\usepackage{times}            %% Choose Times Roman font
\usepackage[english]{babel}   %% This is an English paper
\usepackage[ansinew]{inputenc}%% use 'utf8' or 'ansinew' to support special characters (e.g., umlauts), depends on your editor!
\usepackage[T1]{fontenc}      %% use PS Type1 fonts
\usepackage[sort&compress,numbers]{natbib}	%% bibliography
\usepackage{amsmath,amssymb}
\usepackage{graphicx}
\usepackage[colorlinks=false,pdfborder={0 0 0}]{hyperref}
\usepackage{units}

\usepackage{dirtytalk}
\usepackage{cleveref}
\usepackage[acronym]{glossaries}
\usepackage{makecell}
\usepackage{multirow}
\usepackage{adjustbox}
\usepackage{svg}
\usepackage[skip=0.5\baselineskip]{caption}
\usepackage{subcaption}
\usepackage{hyperref}
\usepackage{booktabs}
\usepackage{tabularx}
\usepackage[skip=0.5\baselineskip]{caption}
\usepackage{calrsfs}
\usepackage{pifont}
\usepackage{balance}

\sloppypar

\newacronym{VAE}{VAE}{Variational Autoencoder}
\newacronym{MI}{MI}{Mutual Information}
\newacronym{fov}{FoV}{factors of variation}
\newacronym{TCVAE}{TCVAE}{Total Correlation Variational Autoencoder}
\newacronym{ELBO}{ELBO}{Evidence Lower Bound}
\newacronym{MIG}{MIG}{Mutual Information Gap}
\newacronym{KLD}{KLD}{Kullback-Leibler divergence}
\newacronym{TC}{TC}{Total Correlation}
\newacronym{gemaps}{GeMAPS}{Geneva minimalistic acoustic parameter set}
\newacronym{LDA}{LDA}{Linear Discriminant Analysis}
\newacronym{lld}{LLD}{low-level descriptor}
\newacronym{FCAE}{FCVAE}{Factorized Convolutational VAE}
\newacronym{FCTCVAE}{TC-FCVAE}{Total Correlation FCVAE}
\newacronym{VTLP}{VTLP}{vocal tract length perturbation}
\newacronym{CPC}{CPC}{Contrastive Predictive Coding}
\newacronym{AWGN}{AWGN}{additive white Gaussian noise}
\newacronym{eer}{EER}{equal error rate}
\newacronym{wer}{WER}{word error rate}

\newcommand{\pthetax}{p_{\boldsymbol{\theta}}(\mathbf{x})}
\newcommand{\qzx}{q_{\boldsymbol{\phi}}(\mathbf{s}\vert{}\mathbf{x})}

\newcommand{\pxz}{p_{\boldsymbol{\theta}}(\mathbf{x}\vert{}\mathbf{s})}

\newcommand{\KL}{\mathrm{KL}}
\newcommand{\latentz}{\mathbf{s}}

\newcommand{\completeness}{Comp.}
\newcommand{\disentanglement}{Mod.}
\newcommand{\explicitness}{Expl.}
\DeclareMathAlphabet{\pazocal}{OMS}{zplm}{m}{n}

\newlength{\texttablesep}
\let\svthefootnote\thefootnote
\newcommand\blankfootnote[1]{%
  \let\thefootnote\relax\footnotetext{#1}%
  \let\thefootnote\svthefootnote%
}

\title{%
    Investigating Speaker Embedding Disentanglement on Natural Read Speech
}

\author{%
    Michael Kuhlmann$^1$,
    Adrian Meise$^1$,
    Fritz Seebauer$^2$,
    Petra Wagner$^2$,
    Reinhold Haeb-Umbach$^1$
}

\address{%
    $^1$Dept. of Communications Engineering, Paderborn University, Email: \texttt{\{kuhlmann,haeb\}@nt.upb.de}, \texttt{atmmeise@mail.upb.de}\\
    $^2$Phonetic Works Group, Bielefeld University, Email: \texttt{\{fritz.seebauer,petra.wagner\}@uni-bielefeld.de}
}

\begin{document}

\maketitle

\begin{abstract}
%% Place your abstract here
Disentanglement is the task of learning representations that identify and separate factors that explain the variation observed in data. Disentangled representations are useful to increase the generalizability, explainability, and fairness of data-driven models. Only little is known about how well such disentanglement works for speech representations. A major challenge when tackling disentanglement for speech representations are the unknown generative factors underlying the speech signal. In this work, we investigate to what degree speech representations encoding speaker identity can be disentangled. To quantify disentanglement, we identify acoustic features that are highly speaker-variant and can serve as proxies for the factors of variation underlying speech. We find that disentanglement of the speaker embedding is limited when trained with standard objectives promoting disentanglement but can be improved over vanilla representation learning to some extent.
\end{abstract}

%% And now start with your paper content

%%%%%%%%%%%%%%%%%%%%%%%%%%%%%%%%%%%%%%%%%%%%%%%%%%%%%%%%%%%%%%%%%%
\blankfootnote{%
    Funded by Deutsche Forschungsgemeinschaft (DFG), project 446378607.
    % and TRR 318/1 2021 - 438445824.
    Computational resources were provided by the Paderborn Center for Parallel Computing.
}
\section{Introduction}
Speech is a rich source of information that conveys not only content-related information but also information about the speaker and the acoustic environment. 
Disentangling those sources of variation would be beneficial for 
a number of downstream tasks, for which either the content-related variations can be considered as noise (e.g., speaker recognition) or the speaker and environment-induced variations (e.g., speech recognition). Further, a fine-grained disentanglement could contribute to phonetic research by providing explicit control over single parameters.

Among the deep generative models, the \gls{VAE}~\cite{vae} is a particularly promising model because it provides an efficient inference mechanism to determine the latent factors of variation from the observed data. Although it has been shown that the vanilla \gls{VAE} is unable to identify the true latent factors of variation from observed data \cite{Khemakhem2020}, this does not render the approach futile. For speech disentanglement where one is interested in separating static from dynamic factors, VAE-based encoder-decoder models with encoders specifically designed for speech have been developed. 
The hierarchical \gls{VAE} \cite{Hsu2017UnsupervisedLO}  and the factorized \cite{fcae} \gls{VAE} use two encoders for unsupervised disentanglement of the speech signal into two disjoint representations where one captures short-term variations and the other long-term variations in the signal.
The former representation, which is generated roughly at frame rate, can be attributed to capturing the linguistic content, and the latter, typically one vector per utterance, represents speaker, style, or environmental factors. We call this latter representation the speaker embedding or encoding.

%Among the deep generative models, the \gls{VAE} appears to be a promising model to achieve those goals as it provides an efficient inference mechanism to determine the latent factors from the observed data. Although it has been shown that the vanilla \gls{VAE} is unable to identify the true latent factors of variation from observed data \cite{Khemakhem2020}, this does not render the approach futile, since in speech, the true latent factors are not known anyway. 
%Indeed, variants of the \gls{VAE}, such as the hierarchical \cite{Hsu2017UnsupervisedLO} and factorized \cite{fcae} \gls{VAE} have been developed that allow the unsupervised disentanglement of the speech signal in variables that capture short-term variations in the signal and other variables that encode long-term variations. The former can be attributed to the linguistic content, and the latter to speaker, style or environmental factors. We call this latter representation the style encoding or style vector in the following.
%Such disentangled representations may show useful in downstream tasks where either the short-term or the long-term variations are of interest while the other can be seen as undesired noise: for speaker or emotion classification the short-term content-induced variations are undesired, while for speech recognition speaker- or environment-induced variations are unwanted. 

In this study, we go one step further and scrutinize the speaker encoding.
Our goal is to reveal the \gls{fov} underlying the speaker encoding and improve disentanglement.
In the following, we refer to this task as (speaker) disentanglement.
Ideally, a disentangled representation encodes all information about one \gls{fov} in one latent dimension (compactness) and the \gls{fov} do not overlap in the latent space (modularity).
This would allow for dedicated manipulation of single factors by latent traversal.
However, most research in this field focuses on analyzing disentanglement on controlled synthetic datasets~\cite{Higgins, Chen, dispeech, locatello2019challenging}.
% TODO: Paper von den Franzosen referenzieren
Here, we tackle disentanglement of the speaker encoding computed from natural speech, whose \gls{fov} are unknown.
We sidestep this issue by taking acoustic features that have a significant impact on speaker identity.

The tools we employ for improving disentanglement are variants of the \gls{VAE}, the 
$\beta$-\gls{VAE}~\cite{Higgins} and the \gls{TCVAE}~\cite{Chen}.
The $\beta$-\gls{VAE} introduces a weighting factor $\beta$ between the two loss terms of the \gls{VAE}, the reconstruction loss and the \gls{KLD} between the variational posterior and a normal isotropic prior, which is a regularization term that punishes deviations from a Gaussian random vector with independent components.
Increasing the weight of the latter improves disentanglement, but at the cost of poorer reconstruction.
The decomposition of the \gls{KLD} proposed by Chen et al.~\cite{Chen} allows a fine-grained control over the different goals the \gls{VAE} objective tries to optimize.

To keep the investigation within the framework of \glspl{VAE} which reconstruct an observation from a latent encoding of the same, we train $\beta$-\glspl{VAE} and \glspl{TCVAE} to disentangle pretrained speaker embeddings (d-vectors) instead of integrating the \gls{VAE} into the speaker embedding training.
To quantify the disentanglement of the speaker embedding, we identify a small set of speaker-variant speech features as proxy factors and measure the modularity and compactness of the speaker embedding dimensions relative to the proxies with the measures defined in~\cite{Eastwood, carbonneau2022measuring}, while informativeness is assessed by speaker recognition experiments.
We find that speaker disentanglement performance is limited and quickly loses informativeness with increasing contribution of the disentanglement objectives.
Yet, we are able to show that these objectives can improve disentanglement to some extent.

%%%%%%%%%%%%%%%%%%%%%%%%%%%%%%%%%%%%%%%%%%%%%%%%%%%%%%%%%%%%%%%%%%
\newacronym{cacomp}{CaComp}{Canonical Completeness}
\newacronym{cadis}{CaDis}{Canonical Disentanglement}

\glsreset{ELBO}

\section{Disentanglement}
% \subsection{$\beta$-VAE and Total Correlation VAE}
The \gls{VAE} was introduced by Kingma and Welling~\cite{vae} as a generative model to approximate the distribution $\pthetax$ underlying a set of observations $\pazocal{X}$ by modeling the joint distribution $p_{\boldsymbol{\theta}}(\mathbf{x}, \latentz)$ between $\pazocal{X}$ and a latent variable $\latentz$, parameterized by $\boldsymbol{\theta}$.
Since the marginal likelihood $\pthetax$ is intractable, an approximate posterior $\qzx$, parameterized by a neural network with parameters $\boldsymbol{\phi}$, is included in the training.
The training criterion for the \gls{VAE} then minimizes the negative approximate marginal log-likelihood
\begin{equation}
    \label{eq:elbo}
    % \log{}\pthetax \approx
    \pazocal{L}_{\operatorname{\beta-VAE}} =
    -\mathbb{E}\left[\log\pxz\right] + \beta_s\mathrm{KL}\left(\qzx\Vert{}p(\mathbf{s})\right),
\end{equation}
where the expectation is taken over $\qzx$ and $\beta_s=1$.
This approximation is known as \gls{ELBO} as it is a lower bound for the true log-likelihood $\log\pthetax$.
The first term in the \gls{ELBO} reflects the reconstruction error $\pazocal{L}_\mathrm{rec}$ where $\boldsymbol{\theta}$ are the parameters of the decoder network which reconstructs the input signal $\mathbf{x}$ from the latent variables $\latentz$.
The second term is a regularization term that encourages the approximate posterior to be close to the prior $p(\latentz)=\pazocal{N}(\latentz;\mathbf{0}, \textbf{I})$.
Here, the approximate posterior is modeled by the encoder which takes feature $\mathbf{x}$ as input and outputs the parameters of the distribution ${q_{\boldsymbol{\phi}}(\latentz|\mathbf{x})=\pazocal{N}\left(\boldsymbol{\mu},\boldsymbol{\sigma}^2\right)}$.
The final latent variable $\latentz$ is then obtained by sampling from 
%the reparameterization trick~\cite{vae} such that 
$q_{\boldsymbol{\phi}}(\latentz|\mathbf{x})$.
% Due to the isotropic Gaussian prior, the \gls{KLD} encourages the latent dimensions in $\latentz$ to be independent of each other.
It is the diagonal covariance structure of the approximate posterior that encourages latent dimensions in $\latentz$ to be independent~\cite{rolinek}.
% If the data $\pazocal{X}$ has generative factors that vary independently from each other, it is likely that different factors are captured in separate dimensions of $\latentz$.

Higgins et al.~\cite{Higgins} showed that setting $\beta_s>1$ results in better disentangled representations.
However, increasing the importance of the regularization runs into the risk of a posterior collapse, i.e., the approximate posterior perfectly follows the prior $p(\latentz)$ but ignores any information from the input $\mathbf{x}$~\cite{lucas2019understanding, lian2022robust}.
% Especially, in speaker-content disentanglement where a second information flow through the content encoder exists, such a posterior collapse might easily happen.
Chen et al.~\cite{Chen} showed that the \glsfirst{KLD} in~\Cref{eq:elbo} can be decomposed into three components:
\begin{align}
    \label{eq:elbo-decomp}
    &\quad \KL(\qzx\Vert{}p(\latentz))
    = \underbrace{\KL\left(q(\latentz ,\mathbf{x})\Vert{}q(\latentz)p(\mathbf{x})\right)}_{\pazocal{L}_\mathrm{MI}} \nonumber \\
    &\quad+\underbrace{\KL\left(q(\latentz)\Vert\prod_jq(s_j)\right)}_{\pazocal{L}_\mathrm{TC}} 
    ~+\underbrace{\sum_j\KL\left(q(s_j)\Vert{}p(s_j)\right)}_{\pazocal{L}_\mathrm{DKL}},
\end{align}
where $s_j$ denotes the $j$-th  dimension of $\latentz$.
The first term refers to the mutual information between observation and latent variables.
The second term measures the \gls{TC}, an extension of the mutual information for more than two random variables, between the dimensions of the latent vector~$\latentz$.
The third term is a dimension-wise \gls{KLD} term to encourage a Gaussian distribution in each latent dimension.
Chen explained the success of the $\beta$-\gls{VAE} with increased importance on the \gls{TC} term.
% On the other hand, as the $\beta$-\gls{VAE} weights all terms equally, decreasing the mutual information may benefit a posterior collapse.
% With the \gls{TCVAE}, we can still encourage disentangled dimensions by properly weighting $\pazocal{L}_\mathrm{TC}$ and simultaneously prevent posterior collapse problems by giving less importance to $\pazocal{L}_\mathrm{MI}$:
The decomposed \gls{ELBO} of the \gls{TCVAE} gives a means to analyze the contribution of the divergence terms for disentanglement by assigning them different importance in the objective
\begin{equation}
    \label{eq:tcvae}
    \pazocal{L}_\mathrm{TCVAE}=
        \pazocal{L}_\mathrm{rec}
        +\alpha\pazocal{L}_\mathrm{MI}
        +\beta\pazocal{L}_\mathrm{TC}
        +\gamma\pazocal{L}_\mathrm{DKL}.
\end{equation}
% Setting $\alpha=\beta=\gamma$ gives the $\beta$-\gls{VAE}.
The $\beta$-VAE is a special case of the \gls{TCVAE} with ${\alpha=\gamma=\beta}$.

\subsection{Quantifying speaker disentanglement}

Quantification of disentanglement is usually done with knowledge of the true \gls{fov}~\cite{carbonneau2022measuring}.
For a complex signal like the speech signal it is difficult to define a reliable set of \gls{fov} that captures all variation.
Therefore, we will start by evaluating the informativeness and separability of the latent space itself.
In a second step, we then propose proxies for the true \gls{fov} in the speech signal~(\Cref{sec:speech-proxies}) which allows us to choose from several supervised metrics.
% the abundant options of supervised disentanglement metrics.

\subsubsection{Unsupervised disentanglement quantification}
A well disentangled latent space should be informative about the observation and the information about the observation stored in different latent dimensions should be separable~\cite{do2020theory}.
We use the following two metrics to measure disentanglement in an unsupervised way. \\
\textbf{\Gls{eer}} on $\mathbf{\hat{x}}$:
If the latent embedding $\mathbf{s}$ is informative about the observation it should allow to faithfully reconstruct $\mathbf{\hat{x}}\sim\pxz$.
To obtain the \gls{eer} we create a custom trial list to measure the cosine similarity between reconstructed observations of same and different speakers and set a threshold such that false alarm and false rejection rates are equal which is the \gls{eer}.
Lower is better. \\
\textbf{WSEPIN}: Do and Tran~\cite{do2020theory} presented an information theoretic metric that measures both separability in the latent space and informativeness w.r.t. the observation:
% WSEPIN is defined as
\begin{equation}
    \label{eq:wsepin}
    \operatorname{WSEPIN} = \sum_j\frac{\rho_j}{H(s_j)}I(\mathbf{x};s_j|\latentz_{\neq{}j}) \geq 0,
\end{equation}
where $I(\mathbf{x};s_j|\mathbf{\latentz}_{\neq{}j}) = I(\mathbf{x};\mathbf{\latentz}) - I(\mathbf{x};\mathbf{\latentz}_{\neq{}_j})$, $I(\mathbf{x};\mathbf{\latentz})$ is the mutual information between $\mathbf{x}$ and $\mathbf{\latentz}$, $H(s_j)$ is the entropy of variable $s_j$, $\rho_j=\frac{I(\mathbf{x};s_j)}{\sum_iI(\mathbf{x};s_i)}$ and $\latentz_{\neq{}j}$ is the latent vector with $s_j$ removed.
If $s_j$ is noisy or encoded in another dimension, then $I(\mathbf{x};\latentz_{\neq{}j})\approx{}I(\mathbf{x};\latentz)$.
On the other hand, if $s_j$ is the only dimension that encodes a particular information about $\mathbf{x}$, i.e., it is separable from $\latentz_{\neq{}j}$ (and informative about $\mathbf{x}$), then $I(\mathbf{x};\latentz_{\neq{}j})<I(\mathbf{x};\latentz)$.
Higher is better.
% Note that we normalize~\Cref{eq:wsepin} by $I(\mathbf{x};\mathbf{\latentz})$ instead of $H(\latentz_j)$ to bound the score between $0$ and $1$.

\subsubsection{Supervised disentanglement quantification}
% When having a factorized representation, we would like to assess the degree of factorization using some quantitative metric (see~\cite{carbonneau2022measuring} for an overview of proposed metrics).
% While there exist unsupervised metrics that work without knowledge of the \gls{fov}, we follow the recommendation from~\cite{carbonneau2022measuring} and measure the degree of factorization with the DCI score~\cite{Eastwood}.
% We discuss the problem of unknown \gls{fov} for speech in \Cref{sec:speech-proxies}.
% A metric that does not rely on balanced \gls{fov} is DCI~\cite{} which measures Disentanglement, Completeness, and Informativeness of a latent embedding for a given set of factors.
We follow the recommendation from~\cite{carbonneau2022measuring} and compute the DCI score~\cite{Eastwood} to check whether the latent dimensions align with the proxy \glspl{fov}.
The DCI score employs the various requirements on a disentangled representation to define a Disentanglement, Completeness, and Informativeness score, where each measures how well a single requirement is fulfilled.
To avoid confusion with the general term \say{disentanglement}, following~\cite{carbonneau2022measuring}, we replace \say{Disenanglement} with \say{modularity}, \say{Completeness} with \say{compactness} and \say{Informativeness} with \say{explicitness}.
To derive these values, separate regressors, one for each \gls{fov}, are trained to predict the factor from the latent variable.
The regressors are chosen such that they allow quantifying the importance $\mathbf{r}_k\in\mathbb{R}^D, k=1,\dots, K,$ of each latent dimension for the regression output, where $K$ denotes the number of \gls{fov} that are evaluated.
\textbf{Compactness (\completeness)} and \textbf{modularity (\disentanglement)} are then calculated as the entropy of the normalized importance scores:
\begin{align}
    \operatorname{\completeness}_k &= 1+\sum_{d=1}^D p_{kd}\log_Dp_{kd},\quad~k=1,\dots, K, \\
    \operatorname{\disentanglement}_d &= 1+\sum_{k=1}^K \tilde{p}_{kd}\log_K\tilde{p}_{kd},\quad~d=1,\dots, D,
\end{align}
where $p_{kd}=r_{kd}/\sum_{d=1}^Dr_{kd}$ and $\tilde{p}_{kd}=r_{kd}/\sum_{k=1}^Kr_{kd}$.
% Compactness captures axis alignment of the latent variable with the factors while modularity measures how many factors are captured in a single dimension of the latent variable.
Both scores are bounded between zero and one, where higher values indicate better performance.
Compactness captures axis alignment of the latent variable with the factors, and it is high if only few dimensions encode information about a factor.
Modularity measures how many factors are encoded in a single dimension of the latent variable and is high if a latent dimension encodes few factors.
Both compactness and modularity simultaneously approach one when latent variables and \gls{fov} form a one-to-one mapping.
%(perfect disentanglement).
% We use disentanglement\_lib~\cite{locatello2019challenging} to compute the compactness and modularity scores.
For \gls{fov} with a continuous domain, explicitness (\explicitness) is measured as the regression error, e.g., mean-squared error, between the predicted and true factor values (lower is better).
% However, since it is difficult to define all \gls{fov} that are relevant for speaker identity, quantifying informativeness from the speaker embedding as defined in~\cite{Eastwood} is unreliable.
% Instead, we report the performance of speaker recognition and speaker verification experiments performed on the speaker embedding as an informativeness score.
% Instead of evaluating informativeness, we report the performance of various downstream tasks on the latent variable as an Informativeness score.
We use disentanglement\_lib~\cite{locatello2019challenging} to compute DCI.

%%%%%%%%%%%%%%%%%%%%%%%%%%%%%%%%%%%%%%%%%%%%%%%%%%%%%%%%%%%%%%%%%%
\section{Identifying proxy factors for speech}
\label{sec:speech-proxies}
% However, any disentanglement metric requires (i) that the \gls{fov} are known such that their influence on each latent dimension can be measured and (ii) that the \gls{fov} are independent.
% TODO: One or two sentences on the generation of natural speech
% For natural speech, the generating factors are unknown and might only be postulated from speech production mechanisms.
% To allow disentanglement measurements for latent representations of natural speech, we use proxy ground truth factors that are extracted from the speech signal.
To employ the DCI metric, we propose to use proxy factors as substitutes for the true unknown \glspl{fov} in a speech signal.
We require these proxies to be highly speaker-variant.
% since we are interested in investigating the disentanglement of the speaker embedding which encodes speaker-variant information.
The set of candidate proxies is the \gls{gemaps} which defines $18$ acoustic \glspl{lld} that have shown to be effective in voice and emotion research~\cite{Eyben}.
We further analyze the set of $N=62$ functionals that are the utterance-wise statistics, e.g., mean and variance, obtained from the \glspl{lld}.
We use openSMILE~\cite{opensmile} to extract the functionals.

To find a good set of proxies explaining speaker variance, we rank the functionals in order of relevance for a speaker recognition task.
We obtain this ranking by performing a \gls{LDA} on the functionals where the class affiliations are given by the speaker identity.
\gls{LDA} finds a projection $\mathbf{W}\in\mathbb{R}^{P\times{}N}$ into a lower dimensional subspace that maximizes the between-class scatter while keeping the within-class scatter constant.
Here, we project the functionals into a one-dimensional subspace $P=1$.
The found projection vector $\mathbf{w}\in\mathbb{R}^N$ will then contain the importance of each functional in discerning different speakers.
The ranking is obtained by sorting the absolute values of the projection vector in descending order.
% A similar analysis with these functionals has been conducted to analyze the acoustic differences between healthy and dysarthric speech~\cite{}.

We apply \gls{LDA} on functionals extracted from the {LibriTTS}~\cite{LibriTTS} train-clean-460 subset which contains utterances from 1100 native English speakers.
$80\%$ of the utterances are used to fit the \gls{LDA} while the remaining $20\%$ are left for testing.
All functionals are normalized by their standard deviation prior to fitting.
% to account for the range differences of the individual functionals.
The test accuracy of the fitted \gls{LDA} is $82.9\%$.
Note that including cepstral features can increase the accuracy to more than $90\%$ but are more difficult to interpret than spectral features.
% and we exclude all percentile functionals resulting in $N=80$ functionals for analysis.
% We expect functionals that are more speaker-variant to have higher importance for correctly recognizing the speakers.
\Cref{tab:functionals-ranking} shows the top-10 ranked functionals which we use as our proxy set (i.e., $K=10$ factors) containing the pitch , first three formant frequencies, slope measures (Hammarberg index and Alpha Ratio) and energy ratios (HNR, H1-A3).

\renewcommand{\arraystretch}{1.2}
\begin{table}[h]
    \centering
    \setlength\tabcolsep{0pt}
    % \resizebox{.48\textwidth}{!}{
        \begin{tabular*}{\linewidth}{@{\extracolsep{1em}}rll}
            \toprule[1.5pt]
             % Rank & \acrshort{gemaps} functional name \\
             % \midrule
             % 1 & F0semitoneFrom27.5Hz\_sma3nz\_amean \\
             % 2 & {HNRdBACF\_sma3nz\_amean} \\
             % 3 & {F3frequency\_sma3nz\_amean} \\
             % 4 & {F1bandwidth\_sma3nz\_amean}  \\
             % 5 & F1frequency\_sma3nz\_amean  \\
             % 6 & hammarbergIndexV\_sma3nz\_amean \\
             % 7 & logRelF0-H1-A3\_sma3nz\_amean \\
             % 8 & F2frequency\_sma3nz\_amean \\
             % 9 & alphaRatioV\_sma3nz\_amean \\
             % 10 & F1frequency\_sma3nz\_stddevNorm \\
             Rank & Functional & Abbr. \\
             \midrule
             1 & pitch (log F0) mean & F0 \\
             2 & Harmonic-to-Noise Ratio (HNR) mean & HNR \\
             3 & 3rd formant frequency (F3) mean & F3 \\
             4 & F1 bandwidth mean & F1bw \\
             5 & 1st formant frequency (F1) mean & F1 \\
             6 & Hammarberg index mean & Hi \\
             7 & \makecell[l]{1st harmonic to F3\\amplitude diff. (H1-A3) mean} & H1-A3 \\
             8 & 2nd formant frequency (F2) mean & F2 \\
             9 & Alpha Ratio mean & $\alpha$\\
             10 & F1 std. dev. & F1std \\
             \bottomrule[1.5pt]
        \end{tabular*}
    % }
    \caption{%
        Top-10 importance ranking of functionals for speaker recognition as found by \gls{LDA}.
    }
    \label{tab:functionals-ranking}
    \vspace{-1em}
\end{table}

%%%%%%%%%%%%%%%%%%%%%%%%%%%%%%%%%%%%%%%%%%%%%%%%%%%%%%%%%%%%%%%%%%
\section{Experiments}
\label{sec:experiments}
Our goal is to disentangle factors representing speaker identity.
As observation space, we use speaker embeddings extracted by a strong d-vector front-end~\cite{cord2023frame}.
The d-vector extractor is a ResNet34 that extracts 256-dimensional speaker embeddings $\mathbf{x}$ (d-vector) from 80-dimensional log-mel spectrograms and was trained on the VoxCeleb speech corpus~\cite{nagrani2020voxceleb}.
This model achieves an \gls{eer} of 0.92\% on a custom speaker verification trial list based on the LibriTTS test-clean subset.
All disentanglement experiments are performed with the LibriTTS speech database.
We use the train-clean-460 subset for training the disentanglement models and the official clean test set for evaluation.
\gls{VAE} encoder and decoder are eight-layered CNNs as in~\cite{fcae}.
The encoder extracts the 128-dimensional latent embedding $\mathbf{s}\sim{}q_\phi(\latentz|\mathbf{x})$ from d-vector $\mathbf{x}$.
All \glspl{VAE} are trained for 100,000 iterations with a learning rate of $10^{-4}$ and batch size of 64.
During training, the weights of the d-vector extractor are fixed and only \gls{VAE} encoder and decoder weights are updated.
\subsection{Unsupervised disentanglement evaluation}
\Cref{tab:beta-vae} shows the \gls{eer} and disentanglement scores for the $\beta$-\gls{VAE} for various $\beta$.
A small value for $\beta$ ($\beta<=10^{-2}$) gives only marginal improvement in measured disentanglement.
When setting $\beta$ higher ($\beta>10^{-2}$), we see a jump in the disentanglement score but at the same time, the speaker encoding will lose too much information about the speaker due to the progression of the posterior collapse (\gls{KLD} approaching zero) and becomes useless for many downstream tasks.
Note that Do and Tran reported values of $\operatorname{WSEPIN}>1$ for well disentangled latent space~\cite{do2020theory}.
For $\beta=0$, we were not able to get a numerically stable estimate of WSEPIN.
\renewcommand{\arraystretch}{1.2}

\begin{table}[h]
    \centering
    % \resizebox{.48\textwidth}{!}{
        \setlength\tabcolsep{0pt}
        \begin{tabular*}{\linewidth}{@{\extracolsep{\fill}}r r r  r@{}}
        \toprule[1.5pt]
        $\alpha=\gamma=\beta$ & \makecell{EER~$\mathbf{\hat{x}}$ [\%]}~$\downarrow$ & WSEPIN~$\uparrow$ & KLD \\
        \midrule
        $0$                &  $\textbf{1.02}$    & NaN & $1731.7$ \\ % noc: /scratch/hpc-prf-nt1/mkuhlmann/storage/speaker_disentanglement/libritts_11
        $10^{-3}$                          &  $1.04$ & $.090$   & $86.6$ \\ % noc:/scratch/hpc-prf-nt1/mkuhlmann/storage/speaker_disentanglement/libritts_21
        $3\times10^{-3}$                          &  $1.98$   & $.118$ & $34.8$ \\ % noc:/scratch/hpc-prf-nt1/mkuhlmann/storage/speaker_disentanglement/libritts_20
        $10^{-2}$                &  $3.59$  & $.150$ & $12.9$ \\ % noc:/scratch/hpc-prf-nt1/mkuhlmann/storage/speaker_disentanglement/libritts_32
        $3\times10^{-2}$                &  $13.10$    & $.290$ & $3.1$ \\ % noc:/scratch/hpc-prf-nt1/mkuhlmann/storage/speaker_disentanglement/libritts_13
        $10^{-1}$                &  $31.54$ & $\mathbf{.345}$ & $2\times10^{-3}$ \\ % noc:/scratch/hpc-prf-nt1/mkuhlmann/storage/speaker_disentanglement/libritts_14
        \bottomrule[1.5pt]
        \end{tabular*}
    % }
    \caption{
        Evaluating disentanglement of the speaker embedding with $\beta$-\gls{VAE} ($\alpha=\gamma=\beta$).
    }
    \label{tab:beta-vae}
    \vspace{-.5em}
\end{table}

\begin{figure*}[t]
    \centering
    \begin{subfigure}[t]{0.46\textwidth}
        \centering
        \includegraphics[width=\textwidth]{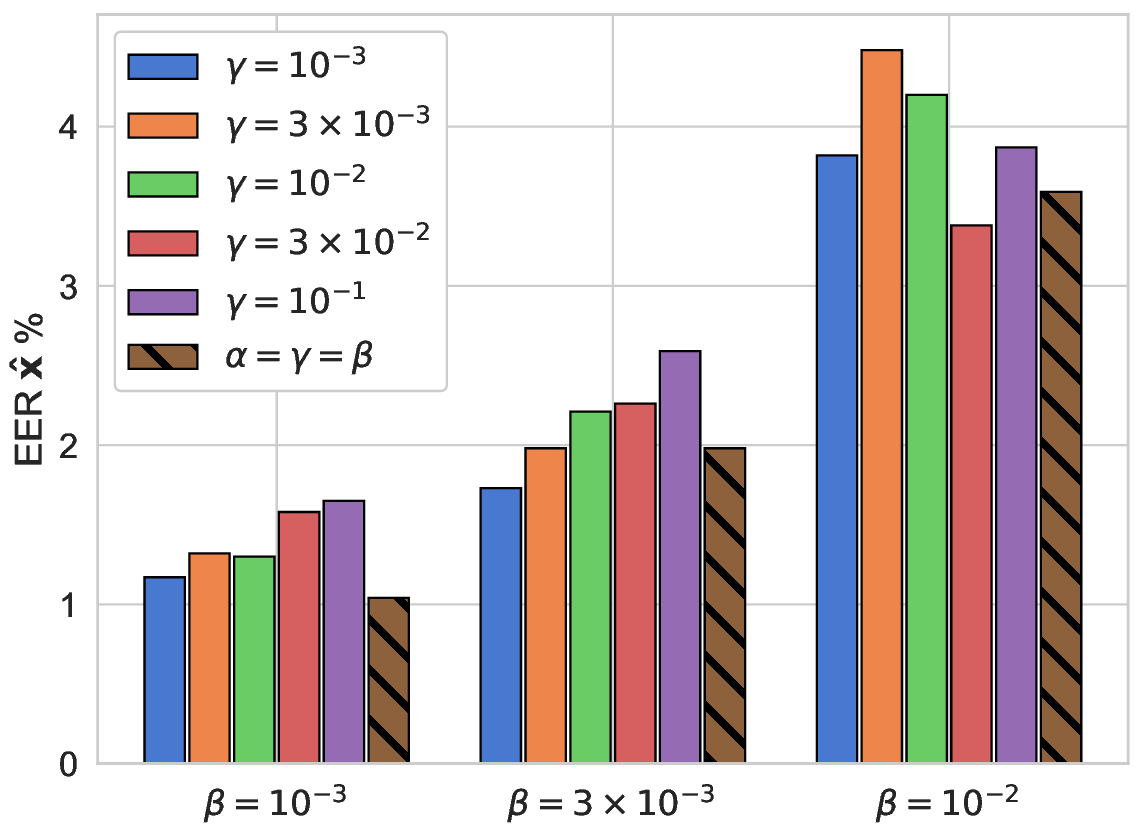}
        % \caption{}
    \end{subfigure}
    \hfill
    \begin{subfigure}[t]{0.46\textwidth}
        \centering
        \includegraphics[width=\textwidth]{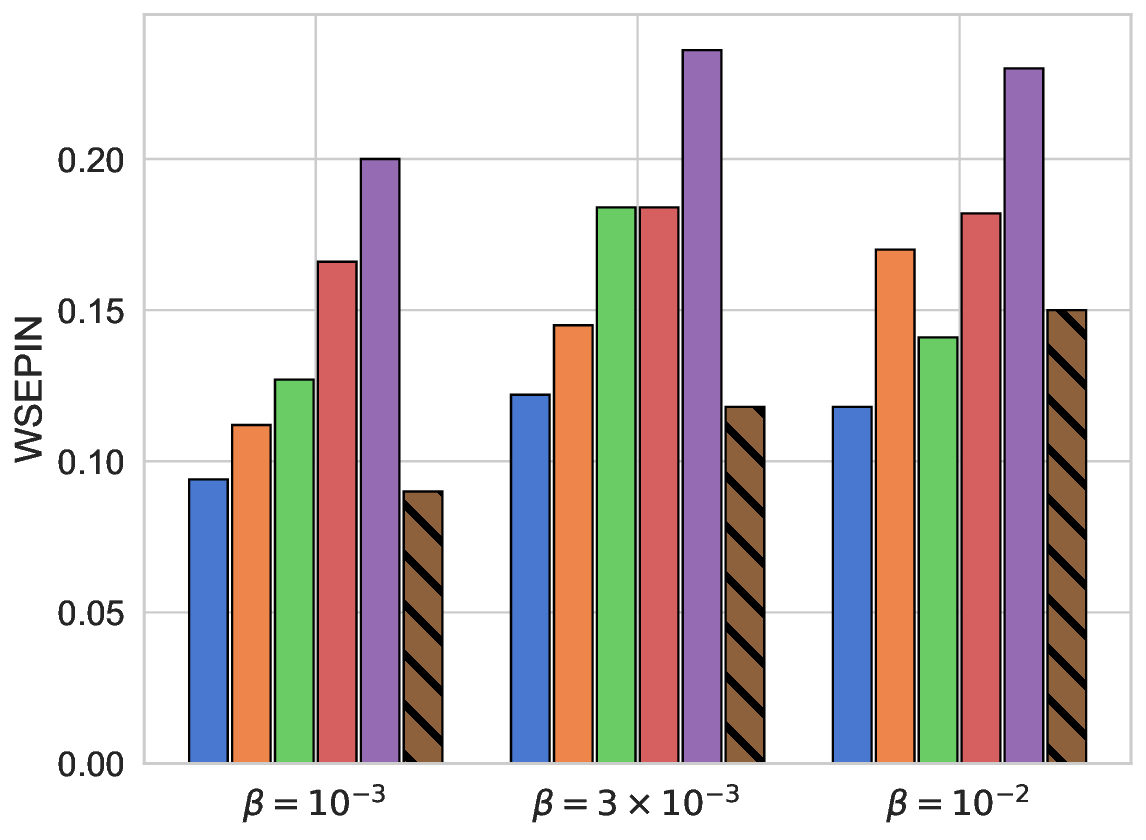}
        % \caption{}
    \end{subfigure}
    \caption{%
        \gls{TCVAE} disentanglement evaluation for various ($\beta$, $\gamma$) sweeps and $\alpha=0$.
        Striped bars ($\alpha=\gamma=\beta$) show the $\beta$-\gls{VAE} results from \Cref{tab:beta-vae} ($\alpha=\beta\neq0$).
        Left: \gls{eer} on reconstructed speaker embedding $\mathbf{\hat{x}}$ (lower is better).
        Right: Separability and informativeness of the latent space as measured by WSEPIN (higher is better).
        Best viewed in color.
        % (a) \Gls{eer} of the reconstructed speaker embedding $\mathbf{\hat{x}}$ (lower is better). \\
        % (b) Disentanglement of the latent space as measured by WSEPIN (higher is better).
    }
    \label{fig:beta-gamma-sweep}
    \vspace{-1em}
\end{figure*}

To further analyze what causes this information loss, we evaluate \glspl{TCVAE} with various combinations of $\beta$ and $\gamma$, keeping $\alpha=0$ fixed to not further penalize mutual information between $\mathbf{x}$ and $\latentz$~(\Cref{fig:beta-gamma-sweep}).
We make the following observations:
% Increasing $\gamma$ while fixing $\beta$ slightly increases \gls{eer} and gives higher WSEPIN.
% Increasing $\beta$ while fixing $\gamma$ gives more drastic increase in \gls{eer}:
% Compared to \inred{ the $\beta$-VAE (?)} the increase in \gls{eer} is less drastic:
Increasing $\gamma$ (fixed $\beta$) by the order of two magnitudes (from $10^{-3}$ to $10^{-1}$) yields a final lower \gls{eer} than increasing $\beta$ (fixed $\gamma$) by the order of a third magnitude.
Also, increasing $\gamma$ while fixing $\beta$ gives consistently better WSEPIN.
A higher value for $\beta$ ($\beta\geq3\times{}10^{-3}$) with a simultaneously high value for $\gamma$ ($\gamma=10^{-1}$) give slightly higher results for separability than $\beta$-\gls{VAE} with same $\beta$ ($\operatorname{WSEPIN}>0.20$) without increasing the \gls{eer} too much ($\operatorname{\gls{eer}}<4\%$).
However, similar to the $\beta$-\gls{VAE}, disentanglement as measured by WSEPIN stays low and a further increase of $\beta$ or $\gamma$ will decrease informativeness.
An explanation for this behavior can be given when the true (unknown) \gls{fov} of the speech signal are assumed to be correlated:
In this case, it is impossible to maximize the data likelihood  and enforce disentanglement of the latent variable at the same time~\cite[Proposition 1]{Trauble}.
We also tried different combinations of $\alpha\neq0$ and $\beta$ while fixing $\gamma=0.1$ and found that \gls{eer} increases and WSEPIN decreases notably with increasing $\alpha$.
\subsection{Supervised disentanglement evaluation}
While WSEPIN allows to rank the models by their degree of disentanglement, it is unclear whether the reached degree is already enough for practical applications.
For example, we would like to be able to alter specific speaker attributes.
Then, it is enough if these attributes exist in disentangled dimensions of the latent encoding.
Here, we evaluate whether acoustic features of the speech signal can be disentangled by the \glspl{VAE}.
\Cref{tab:dci} shows the DCI scores evaluated with the proxy factors for various \glspl{VAE}.
\renewcommand{\arraystretch}{1.2}

\begin{table}[h]
    \centering
    \setlength\tabcolsep{0pt}
    \begin{tabular*}{\linewidth}{@{\extracolsep{\fill}}r r l r r r@{}}
    \toprule[1.5pt]
    $\beta$ & $\alpha$ & $\gamma$ & \completeness$\uparrow$ & \disentanglement$\uparrow$ & \explicitness$\downarrow$ \\
    \midrule
    0          & $\beta$ & $\beta$ &  .37     &  $\mathbf{.37}$    &  .27    \\
    $10^{-3}$  & $\beta$ & $\beta$ &  .44     &   .20   &  .27    \\
    $3\times{}10^{-3}$ & $\beta$ & $\beta$ &   .43     &   .26   &  .27    \\
    $10^{-2}$ & $\beta$ & $\beta$ &    $\mathbf{.45}$     &   .21   &  .28    \\
    $3\times{}10^{-2}$ & $\beta$ & $\beta$  &   .44    &  .20   & .30     \\
    $10^{-1}$ & $\beta$ & $\beta$ &   .27     &   .25   &  .31    \\
    \midrule
    $10^{-3}$ & 0 & $10^{-1}$  &   .31   &   $\mathbf{.26}$   &   .27 \\ 
    $3\times10^{-3}$ & 0 & $10^{-1}$ & .34 & .24 & .28 \\
    $10^{-2}$ & 0 & $10^{-1}$ & $\mathbf{.48}$ & .23 & .28 \\
    \bottomrule[1.5pt]
    \end{tabular*}
    \caption{%
        DCI scores with proxy factors for various weight combinations obtained from the latent embedding $\latentz$.
    }
    \label{tab:dci}
    \vspace{-.5em}
\end{table}

\begin{figure}[h]
    \centering
    \begin{subfigure}[h]{\linewidth}
        \centering
        \includegraphics[width=\textwidth]{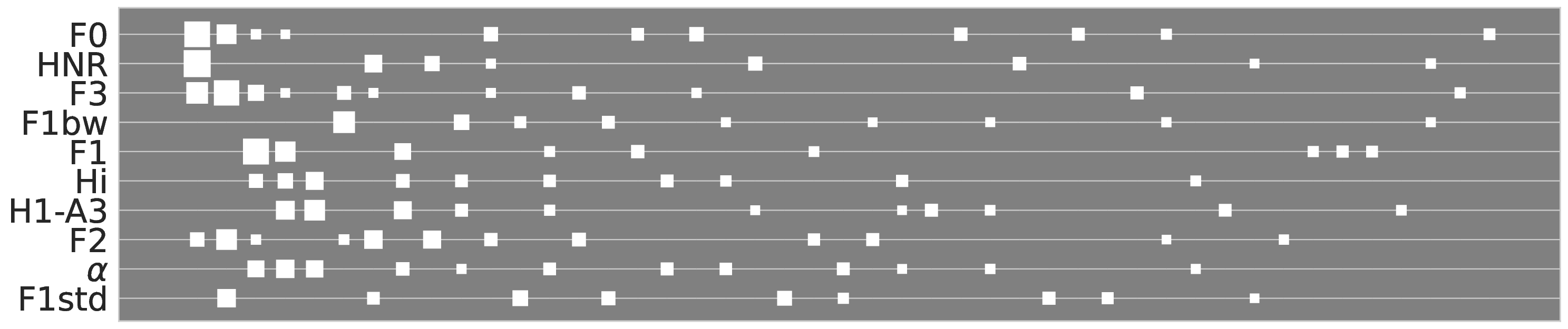}
        \caption{$\mathbf{x}$}
    \end{subfigure}
    \hfill
    \begin{subfigure}[h]{\linewidth}
        \centering
        \includegraphics[width=\textwidth]{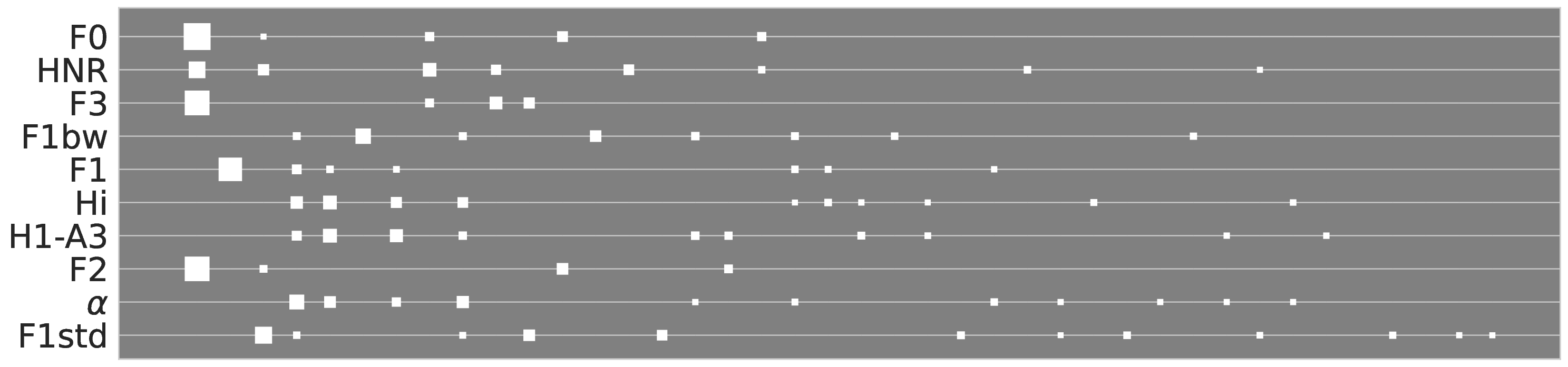}
        \caption{$\latentz$ ($\alpha=\beta=\gamma=0$)}
    \end{subfigure}
    \hfill
    \begin{subfigure}[h]{\linewidth}
        \centering
        \includegraphics[width=\textwidth]{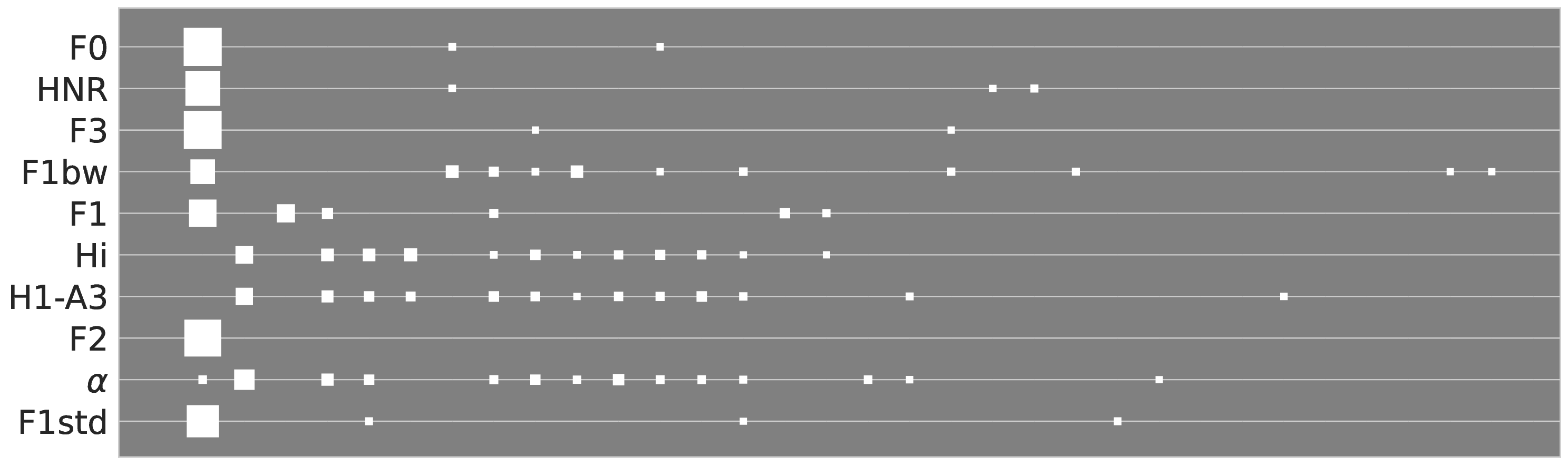}
        \caption{%
            $\latentz$ from \gls{TCVAE} with $\alpha=0$, $\beta=10^{-2}$ and $\gamma=10^{-1}$
        }
    \end{subfigure}
    \caption{%
        Hinton diagram: (Proxy) \gls{fov} over embedding dimensions.
        Larger squares denote higher importance for predicting a factor.
        % Factors are sorted by their speaker recognition importance.
        \gls{fov} abbreviations taken from \Cref{tab:functionals-ranking}.
    }
    \label{fig:hinton}
    \vspace{-2em}
\end{figure}

As a baseline, we also compute the DCI score from observation $\mathbf{x}$ which gives a compactness of $0.29$, modularity of $0.36$ and explicitness of $0.26$.
The autoencoder ($\beta=0$) achieves the same degree of modularity but higher compactness due to the lower dimensionality of the latent space.
Increasing the regularization on the latent space gives better compactness except for weight combinations where both $\alpha$ and $\beta$ are low.
Unfortunately, modularity decreases, i.e., a latent dimension will encode more proxy factors.
For practical applications, a high modularity is desirable as it allows to change a single factor by manipulating the latent dimensions encoding it~\cite{carbonneau2022measuring}.
However, the results for DCI (and any other supervised metric) depend on the appropriate choice of proxy factors.
The results here show that the selected acoustic features are not the factors the \gls{VAE} chooses to disentangle.
This is reasonable as the proxies we have used are highly correlated.
\Cref{fig:hinton} shows the Hinton diagrams~\cite{Eastwood} visualizing the importance matrix of the regressor.
Compared to the observation and latent space of the autoencoder, the \gls{TCVAE} needs less dimensions to encode information about the factors:
The first dimension encodes most of the information about the pitch and formants (higher excitation, i.e., pitch, benefits higher resonants, i.e., formants) while the second dimension encodes slope measures.
% For this particular model, the third dimension encodes only (partial) information about F1 (modularity=1) which may allow for F1 manipulation.
%

%%%%%%%%%%%%%%%%%%%%%%%%%%%%%%%%%%%%%%%%%%%%%%%%%%%%%%%%%%
\section{Conclusions}
Disentanglement of a speaker embedding extracted from natural speech is a challenging task due to the unknown factors of variation in speech.
To measure the disentanglement of the speaker embedding, we relied on a small set of acoustic features that we identified to be highly speaker-variant.
We found that speaker disentanglement can be improved to a certain degree using common disentanglement approaches and mainly improved in compactness of the latent space.
Enforcing the factorization too strictly quickly resulted in uninformative speaker embeddings.
Therefore, future work should develop dedicated strategies to disentangle specific attributes of the speech signal, e.g.,~\cite{sadok2023learning}.
% To further improve the disentanglement of speaker embeddings, future work should develop dedicated strategies 
% To further improve the factorization of speaker embeddings, future work should develop strategies on how to deal with the high correlations between the generative factors underlying the speech signal.

%%%%%%%%%%%%%%%%%%%%%%%%%%%%%%%%%%%%%%%%%%%%%%%%%%%%%%%%%%

% \vspace{1em}
% \noindent\textbf{Acknowledgements}\\
% This research was funded by Deutsche Forschungsgemeinschaft (DFG), projects 446378607 and TRR 318/1 2021 - 438445824.
% Computational resources were provided by the Paderborn Center for Parallel Computing.

% BIBLIOGRAPHY
%%%%%%%%%%%%%%%%%%%%%%%%%%%%%%%%%%%%%%%%%%%%%%%%%%%%%%%%%%%%%%%%%%
% \balance
\bibliographystyle{ieeetr}
\bibliography{mybib}

%%%%%%%%%%%%%%%%%%%%%%%%%%%%%%%%%%%%%%%%%%%%%%%%%%%%%%%%%%%%%%%%%%

\end{document}